\documentclass[aps,prl,reprint,superscriptaddress]{revtex4-2}

\usepackage[utf8]{inputenc}
\usepackage[T1]{fontenc}
\usepackage{graphicx}
\graphicspath{{figure/}}
\usepackage{bm}
\usepackage{amsmath,amssymb}
\usepackage{xcolor}
\usepackage[
colorlinks=true,
citecolor=blue,
urlcolor=blue,
linkcolor=blue
]{hyperref}

\begin{document}
	
	\title{Displacement-field-driven reconstruction of low energy transport in
		few-layer PtSe$_2$}
	
	\author{Xiao Liu}
	\affiliation{Department of Materials Science and Engineering, National
		University of Singapore, Singapore 117575, Singapore}
	
	\author{Yaroslav Zhumagulov}
	\affiliation{Institute of Physics, Ecole Polytechnique F\'ed\'erale de Lausanne
		(EPFL), CH-1015 Lausanne, Switzerland}
	
	\author{Yuang Jie}
	\affiliation{Department of Materials Science and Engineering, National
		University of Singapore, Singapore 117575, Singapore}
	
	\author{Ahmet Enes Bozcali}
	\affiliation{Department of Materials Science and Engineering, National
		University of Singapore, Singapore 117575, Singapore}
	
	\author{Johan F\'elisaz}
	\affiliation{Institute of Physics, Ecole Polytechnique F\'ed\'erale de Lausanne
		(EPFL), CH-1015 Lausanne, Switzerland}
	
	\author{Qi Zhang}
	\affiliation{Department of Materials Science and Engineering, National
		University of Singapore, Singapore 117575, Singapore}
	
\author{Old\v{r}ich Cicv\'arek}
	\affiliation{Department of Inorganic Chemistry, University of Chemistry and
		Technology Prague, Technick\'a 5, 166 28 Prague 6, Czech Republic}
	
	\author{Kenji Watanabe}
	\affiliation{Research Center for Electronic and Optical Materials, National
		Institute for Materials Science, 1-1 Namiki, Tsukuba 305-0044, Japan}
	
	\author{Takashi Taniguchi}
	\affiliation{Research Center for Materials Nanoarchitectonics, National
		Institute for Materials Science, 1-1 Namiki, Tsukuba 305-0044, Japan}
	
	\author{Zden\v{e}k Sofer}
	\affiliation{Department of Inorganic Chemistry, University of Chemistry and
		Technology Prague, Technick\'a 5, 166 28 Prague 6, Czech Republic}
	
	\author{Oleg V. Yazyev}
	\affiliation{Institute of Physics, Ecole Polytechnique F\'ed\'erale de Lausanne
		(EPFL), CH-1015 Lausanne, Switzerland}
	
	\author{Ahmet Avsar}
	\email{aavsar@nus.edu.sg}
	\affiliation{Department of Materials Science and Engineering, National
		University of Singapore, Singapore 117575, Singapore}
	\affiliation{Department of Physics, National University of Singapore, Singapore
		117542, Singapore}
	\affiliation{Centre for Advanced 2D Materials, National University of
		Singapore, Singapore 117546, Singapore}
	
	\date{\today}
	
	\begin{abstract}
		In layered semiconductors, a perpendicular displacement field generates an
		interlayer potential difference that competes with interlayer hybridization,
		modifying both the band gap and the finite-density electronic states that carry
		current. Resolving this interplay requires a material lying close to the
		semiconductor-to-semimetal transition, where moderate electric fields can
		strongly reshape the low-energy electronic structure. Here, we investigate
		displacement-field-driven transport in dual-gated semiconducting PtSe$_2$,
		whose pronounced thickness-dependent electronic structure provides access to
		this low-band gap regime. Unlike thinner layers, the displacement-field
		response is strong in six-layer PtSe$_2$, which lies at the verge of the
		semiconductor-to-semimetal crossover with only a small residual transport gap.
		Even weak displacement fields rapidly suppress this residual gap near charge
		neutrality, driving the system toward a band-overlap regime. At the same time,
		the conductivity decreases in the heavily hole-doped regime, demonstrating that
		the displacement field modifies not only the gap but also the conducting
		valence-band states. Fixed-relaxation-time Wannier transport calculations
		reproduce both responses, showing that they originate from field-induced band
		overlap together with reconstruction of the valence-band dispersion. These
		results establish finite-density transport as a sensitive probe of
		displacement-field-driven electronic structure reconstruction and extend electrical
		control beyond conventional band-gap engineering.
	\end{abstract}
	
	\maketitle

	\begin{figure*}[t]
		\centering
		\includegraphics[width=0.95\textwidth]{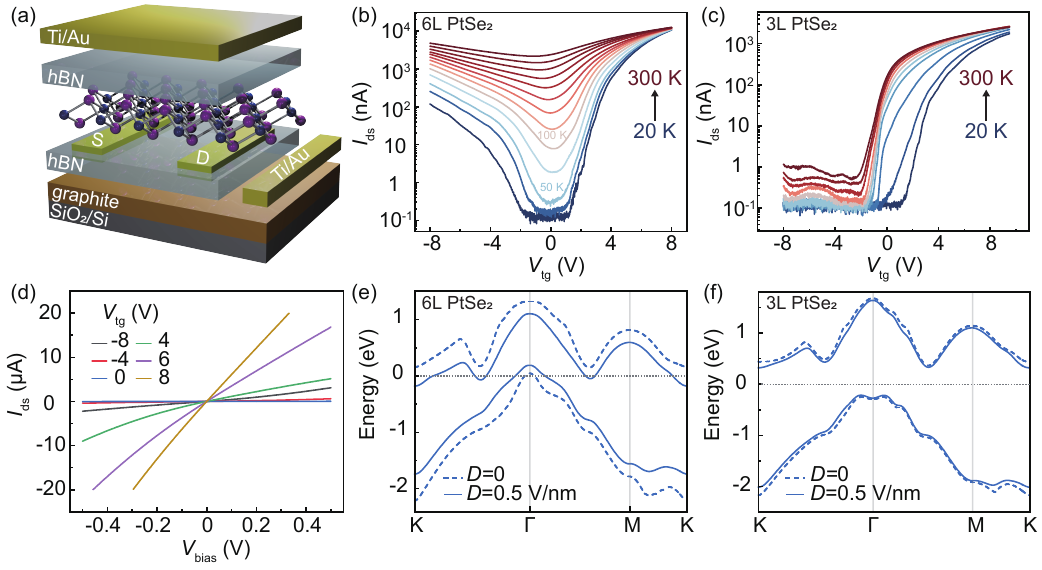}
		\caption{Device structure and basic transport characteristics of six-layer
			PtSe$_2$. (a) Schematic of the dual-gated few-layer PtSe$_2$ device. The
			PtSe$_2$ channel is placed on pre-patterned Pt contact electrodes and
			encapsulated by top and bottom hBN layers. A Ti/Au top gate and a graphite
			bottom gate enable dual-gate control of carrier density and out-of-plane
			displacement field. (b) and (c) Temperature-dependent transfer
			characteristics of the six- and three-layer devices, respectively, measured
			from 20 to 300 K at $V_{\mathrm{bg}}=0$ V and
			$V_{\mathrm{bias}}=0.2$ V. (d) Output characteristics measured at 100 K and
			$V_{\mathrm{bg}}=0$ V for different $V_{\mathrm{tg}}$ values. (e) and (f)
			Schematic low-energy band structures of six- and three-layer PtSe$_2$,
			respectively. The displacement field drives six-layer PtSe$_2$ toward band
			overlap, whereas the corresponding band-structure change is much weaker in
			three-layer PtSe$_2$.}
		\label{fig:device}
	\end{figure*}
	
	
	Electric fields provide a powerful means to manipulate electronic band
	structures through the Stark
	effect~\cite{miller1984bandedge,miller1985electric}. Layered two-dimensional
	materials are particularly well suited for this purpose because their
	atomic-scale thickness and weak out-of-plane screening allow a perpendicular
	displacement field to establish a substantial interlayer potential difference,
	thereby shifting the relative energies of electronic states with different
	layer contributions~\cite{chaves2020bandgap,
		castellanosgomez2013screening,tian2020polarizability}. This enables electric
	fields to compete directly with interlayer hybridization and reconstruct the
	low-energy electronic structure~\cite{min2007bilayer,
		mccann2006asymmetry,oostinga2008gate,zhang2009tunable}. Such
	displacement-field engineering has been demonstrated in several layered
	semiconductors, including few-layer black
	phosphorus~\cite{kim2015blackphosphorus,deng2017blackphosphorus,
		liu2017giantstark}, multilayer
	WSe\textsubscript{2}~\cite{dai2015wse2,domaretskiy2022quenching}, and
	MoS\textsubscript{2}~\cite{liu2012bilayermos2,chu2015tunable}, where vertical
	electric fields substantially reduce the band gap and, in some cases, drive gap
	closure and a semiconductor-to-semimetal transition. These studies demonstrate
	strong Stark-effect band-gap tuning in layered materials. However, the same
	field-induced band modulation is expected to modify the band-edge states that
	carry current~\cite{tayari2016velocity,masseroni2021mos2}, while its
	consequences for finite-density transport remain largely unexplored.
	
	PtSe\textsubscript{2} is a layered transition-metal dichalcogenide with a
	strongly thickness-dependent electronic structure. As the thickness increases,
	its band gap rapidly decreases from the semiconducting monolayer and bilayer
	limits, with reported gaps of \textasciitilde1.85 eV and \textasciitilde0.75
	eV, respectively, toward a semimetallic state in thicker
	samples~\cite{yan2017ptse2,zhao2017ptse2,li2021layer,zhang2021precise,
		ansari2019quantum,villaos2019thickness,ciarrocchi2018ptse2,
		wang2015monolayer,mejamai2026mit}. This evolution is associated with reduced
	quantum confinement and an increasing contribution of interlayer coupling to
	the low-energy electronic structure. Interlayer hybridization of the
	out-of-plane Se \emph{p}\textsubscript{z} orbitals splits the corresponding
	valence states into bonding and antibonding components, with the antibonding
	state shifting upward and narrowing the bandgap~\cite{zhang2017mechanism}. As
	the intrinsic gap becomes smaller, experimentally accessible displacement
	fields become comparable to the energy scales governing the low-energy
	electronic structure, making few-layer PtSe\textsubscript{2} an ideal platform
	for studying electric-field-driven electronic reconstruction.
	
	Here, we investigate displacement-field-driven transport in dual-gated
	few-layer PtSe\textsubscript{2} across its thickness-dependent semiconducting
	regime. The field response is weak in thinner, larger-gap crystals but becomes
	pronounced in six-layer PtSe\textsubscript{2}, which lies near the
	semiconductor-to-semimetal crossover with only a small residual transport gap.
	By combining transport measurements with Wannier-based conductance
	calculations, we reveal two distinct signatures of electric-field-driven
	electronic reconstruction: rapid suppression of the residual transport gap near
	charge neutrality and a simultaneous reduction of conductivity in the heavily
	hole-doped regime arising from reconstruction of the valence-band dispersion.
	The ability to electrostatically reconstruct the conducting bands could enable
	access to electronic states and phases absent from the unperturbed band
	structure.
	
	
	The dual-gated few-layer PtSe\textsubscript{2} field-effect transistor used in
	this work is schematically illustrated in Fig. 1(a). The
	PtSe\textsubscript{2} channel is fully encapsulated by hexagonal boron nitride
	(hBN) and contacted by Pt electrodes, while few-layer graphite and Ti/Au serve
	as the back gate and top gate, respectively. This dual-gate geometry enables
	independent control of the carrier density and the vertical electric field in
	the PtSe\textsubscript{2} channel. Devices with three-, five-, and six-layer
	PtSe\textsubscript{2} channels were investigated to examine the thickness
	dependence of the displacement-field response. Unless otherwise stated, the
	results discussed below were obtained from the six-layer
	PtSe\textsubscript{2} device.
	
	Transfer characteristics of six-layer PtSe\textsubscript{2} were measured from
	20 to 300 K by sweeping the top-gate voltage
	(\emph{V}\textsubscript{tg}) at zero back-gate voltage
	(\emph{V}\textsubscript{bg} = 0 V), with the top-gate leakage current kept
	below 50 pA throughout the measurements (Fig. 1b). The temperature dependence
	of the transfer curves reflects semiconducting transport in six-layer
	PtSe\textsubscript{2}. Near charge neutrality, the Fermi level lies in the band
	gap and the carrier density is strongly suppressed at low temperature. As the
	temperature increases, thermally activated carriers populate states near the
	band edges, increasing the minimum current and narrowing the off-state region.
	Notably, the minimum current rises rapidly between 20 and 100 K, with the
	low-current state already strongly suppressed in the 50-100 K range. This
	strong thermal sensitivity indicates that only a small residual transport gap
	remains in six-layer PtSe\textsubscript{2}, placing it close to the
	semiconductor-to-semimetal crossover. At 100 K, the drain current-bias voltage
	characteristics are nearly linear over the investigated gate-voltage range
	[Fig. 1(d)], indicating efficient carrier injection under these conditions.
	The 50-100 K range therefore provides a particularly sensitive regime for
	examining how a displacement field perturbs this residual gap and the associated
	conducting states, and will be the primary temperature range considered below.
	Away from charge neutrality, the on-state current also increases with
	temperature, most prominently in the heavily hole-doped regime, indicating that
	thermally activated processes, such as carrier injection across contact
	barriers or thermal activation from band-tail states, contribute to the measured
	conductance~\cite{liu2016vdw,kang2014contacts}. The temperature dependence is
	considerably weaker on the electron-doped side at large positive
	\emph{V}\textsubscript{tg} than on the hole-doped side at large negative
	\emph{V}\textsubscript{tg}.
		\begin{figure}[!t]
		\centering
		\includegraphics[width=\columnwidth]{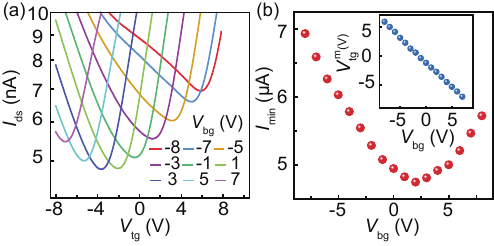}
		\caption{Back-gate modulation of the transfer characteristics at 300 K.
			(a) Transfer characteristics measured at 300 K under different
			$V_{\mathrm{bg}}$, with $V_{\mathrm{bias}}=0.5$ V.
			(b) Minimum current $I_{\min}$ extracted from the transfer curves in (a) as
			a function of $V_{\mathrm{bg}}$. Inset: top-gate voltage at the
			minimum-current point, denoted $V_{\mathrm{tg}}^{m}$, plotted as a function
			of $V_{\mathrm{bg}}$.}
		\label{fig:backgate}
	\end{figure}
	
	The larger-gap regime is illustrated by the three-layer device measured under
	the same conditions [Fig. 1(c)]. In contrast to six-layer
	PtSe\textsubscript{2}, the low-current region extends over a substantially
	wider top-gate voltage range, and hole conduction is not reached within the
	accessible gate-voltage window. The off-state current also remains below
	approximately 1 nA up to room temperature. The extended low-current region and
	its weaker thermal evolution are consistent with the substantially larger band
	gap of three-layer PtSe\textsubscript{2}, as also seen in the calculated
	zero-field band structures [Figs. 1(e) and 1(f)]. The calculations further show
	how this difference in the intrinsic gap determines the response to a
	perpendicular electric field. In six-layer PtSe\textsubscript{2}, the field
	strongly reduces the band separation and drives the system toward band overlap,
	whereas the larger-gap three-layer PtSe\textsubscript{2} is only weakly
	affected over the same field range. This pronounced sensitivity of the low-gap
	six-layer regime motivates our focus below on its displacement-field-dependent
	transport; corresponding measurements on three- and five-layer devices are
	presented in the Supplemental Material~\cite{supplemental}.
	
	We next examine the dual-gate response of six-layer PtSe\textsubscript{2},
	beginning at 300 K. The transfer characteristics depend strongly on
	\emph{V}\textsubscript{bg}, as shown by the
	\emph{I}\textsubscript{ds}-\emph{V}\textsubscript{tg} curves in Fig. 2(a). As
	\emph{V}\textsubscript{tg} is swept, the device is continuously tuned from
	n-type to p-type conduction through the charge-neutrality point (CNP), while
	systematically shifts the CNP and modifies the minimum current. The extracted
	minimum-current values are summarized in Fig.2(b). The inset of Fig.2(b) shows
$V_{\mathrm{tg}}^{m}$, defined as the
	\emph{V}\textsubscript{tg} at the minimum-current point, as a function of
	\emph{V}\textsubscript{bg}. The linear
$V_{\mathrm{tg}}^{m}$--\emph{V}\textsubscript{bg}
	relation indicates effective capacitive compensation between the top and back
	gates, allowing the carrier density \emph{n} and the displacement field
	\emph{D} to be independently controlled in the following measurements.

	\begin{figure*}[!t]
		\centering
		\includegraphics[width=\textwidth]{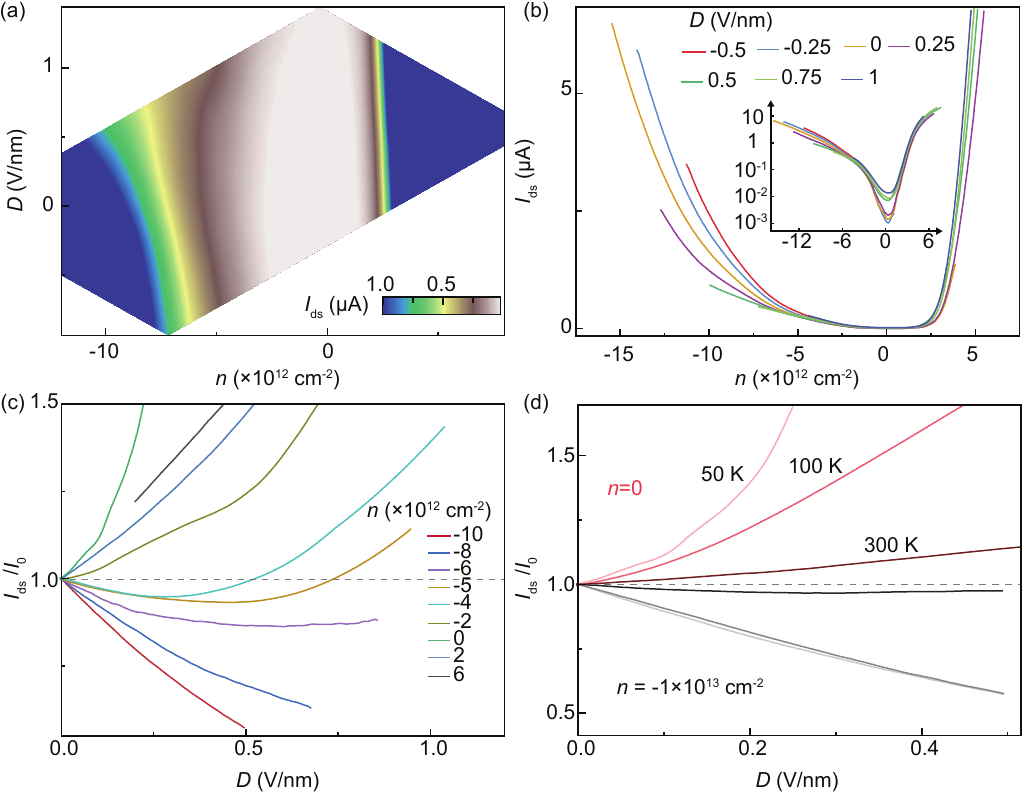}
		\caption{Displacement-field modulation of transport in six-layer PtSe$_2$.
			(a) Current map measured at 50 K as a function of carrier density $n$ and
			displacement field $D$, with $V_{\mathrm{bias}}=0.5$ V. With increasing
			$D$, the current near charge neutrality increases. (b) Horizontal line cuts
			from the map in (a), showing the current as a function of $n$ at selected
			$D$ values. Inset: semilog plot zoomed in near the CNP. (c) Vertical line
			cuts from the map in (a), plotted as normalized current
			$I_{\mathrm{ds}}/I_0$ as a function of $D$ at selected carrier densities.
			Here $I_0$ denotes the current at $D=0$ for each fixed $n$, either measured
			directly or obtained by extrapolation. The dashed line marks
			$I_{\mathrm{ds}}/I_0=1$. (d) Normalized current $I_{\mathrm{ds}}/I_0$
			plotted as a function of $D$ at fixed carrier densities $n=0$ (red-toned
			curves) and $n=-1\times10^{13}$ cm$^{-2}$ (gray-toned curves) for 50, 100,
			and 300 K. For each density, lighter colors denote lower temperatures, and
			darker colors denote higher temperatures.}
		\label{fig:transport-map}
	\end{figure*}

	At 50 K, where the residual transport gap remains clearly resolved, the
	displacement-field response becomes pronounced. Increasing \emph{D} rapidly
	narrows the low-current region around charge neutrality in the
	\emph{n}-\emph{D} map [Fig. 3(a)]. The corresponding line cuts [Fig. 3(b)]
	show that even modest displacement fields strongly enhance the minimum current,
	indicating rapid suppression of the residual transport gap. This behavior is
	consistent with the strong field-induced gap reduction predicted for six-layer
	PtSe\textsubscript{2} in Fig. 1(e). Away from charge neutrality, however, the
	displacement-field response changes qualitatively. To resolve this evolution,
	we take vertical cuts through the \emph{n}-\emph{D} map at fixed carrier
	density and plot the normalized current
	\emph{I}\textsubscript{ds}/\emph{I}\textsubscript{0}, where
	\emph{I}\textsubscript{0} is the current at \emph{D}=0 [Fig. 3(c)]. Only the
	\emph{D} \textgreater{} 0 data are shown here; the corresponding
	\emph{D} \textless{} 0 response is presented in Fig. S4(a). The strongest
	current enhancement occurs near the CNP, while on the electron-doped side the
	current increases monotonically with \emph{D}. On the hole-doped side,
	however, the response evolves strongly with carrier density:
	\emph{I}\textsubscript{ds}/\emph{I}\textsubscript{0} increases with
	\emph{D} in the lightly hole-doped regime, becomes nonmonotonic at intermediate
	densities, and decreases with increasing \emph{D} in the heavily hole-doped
	regime. The reversal of the field response deep in the valence band cannot be
	explained by suppression of the transport gap alone and instead points to
	field-induced reconstruction of the conducting states.

	To connect these observations to the underlying electronic structure, we
	evaluate the Wannier tight-binding model conductance of six-layer PtSe\textsubscript{2} under
	a displacement field [Fig. 4(a)]. Around charge neutrality, the calculated
	low-conductance region is progressively suppressed with increasing field.
	Conductance line cuts at selected displacement fields [Fig. 4(b)] reproduce
	the main experimental trend: the conductance near charge neutrality increases
	with \emph{D} and becomes much less dependent on carrier density at high
	fields, consistent with strong suppression of the low-energy gap and the
	approach toward band overlap. Crucially, the calculation also reproduces the
	opposite response deep in the hole-doped regime. As shown by the fixed-density
	line cuts in Fig. 4(c), the conductance decreases with increasing displacement
	field at high hole density, in agreement with the measured suppression. Because
	the calculation assumes a fixed relaxation time and includes neither contact
	resistance nor field-dependent disorder scattering, this suppression arises
	from the field-induced evolution of the band structure itself. The displacement
	field reconstructs the low-energy valence-band states, modifying their
	dispersion, band velocities, and Fermi-contour geometry rather than simply
	changing the number of available states at the Fermi level. At lower hole
	densities, by contrast, field-induced gap reduction and the approach toward
	band overlap enhance conduction, while the nonmonotonic response at
	intermediate densities reflects a crossover between these competing
	contributions. Wannier tight-binding model calculations capture the intrinsic, field-even
	conductance response, while an additional polarity-dependent contribution
	observed experimentally is associated with the asymmetric device environment
	(Supplemental Material~\cite{supplemental}).

	\begin{figure*}[!t]
	\centering
	\includegraphics[width=\textwidth]{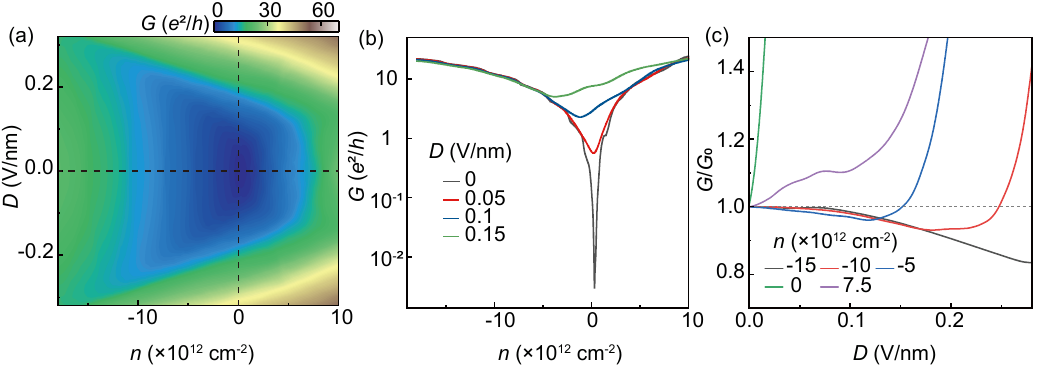}
		\caption{Wannier-model conductance of six-layer PtSe$_2$ under a displacement
			field. (a) Calculated longitudinal conductance $G$ of six-layer PtSe$_2$ as
			a function of carrier density $n$ and displacement field $D$. $G$ is
			normalized by $e^2/h$, where $e$ is the elementary charge and $h$ is
			Planck's constant. The dashed lines mark charge neutrality and zero field.
			(b) Conductance as a function of carrier density at selected displacement
			fields. (c) Normalized conductance $G/G_0$ as a function of displacement
			field at selected carrier densities, where $G_0$ is the conductance at zero
			displacement field for the same carrier density.}
		\label{fig:wannier}
	\end{figure*}
	
	Finally, the displacement-field response weakens with increasing temperature
	[Fig. 3(d)]. At the CNP, the current enhancement decreases progressively upon
	warming, consistent with the thermal suppression of the residual
	low-conductance state already evident in Fig. 1(b). In the heavily hole-doped
	regime, the field-induced conductivity suppression remains clearly resolved at
	50 and 100 K but becomes much weaker at 300 K, where
	\emph{I}\textsubscript{ds}/\emph{I}\textsubscript{0} remains close to unity
	over the measured field range. The strongest signatures of
	displacement-field-driven reconstruction therefore occur in the same
	50-100 K regime identified from Fig. 1, where the residual low-conductance
	state is already fragile and the electronic structure remains highly
	susceptible to electric-field perturbation.
	
	
	Our results identify a displacement-field tuning regime distinct from that of
	other layered systems, where a perpendicular field either opens a gap in a
	gapless band structure, as in bilayer graphene
	~\cite{min2007bilayer,mccann2006asymmetry,oostinga2008gate,zhang2009tunable},
	or suppresses the gap, as in multilayer BP, WSe\textsubscript{2} and
	MoS\textsubscript{2}
	~\cite{kim2015blackphosphorus,deng2017blackphosphorus,
		liu2017giantstark,dai2015wse2,domaretskiy2022quenching,
		liu2012bilayermos2,chu2015tunable,masseroni2021mos2}. In six-layer
	PtSe\textsubscript{2}, the field tunes the band structure and thereby produces
	two effects on transport. Near charge neutrality, it drives the conduction band
	and valence band edges toward each other and eventually into overlap, which
	raises the minimum conductance. At large hole densities, where the Fermi level
	lies well inside the valence band, a change of the gap alone leaves the states
	at the Fermi level unaffected. The observed conductance suppression therefore
	requires a modification of the valence band dispersion. We also find that a
	small gap is necessary but not sufficient for strong field control. The
	response depends on whether the interlayer potential difference built up across
	the channel, which is limited by the channel thickness and internal dielectric
	screening ~\cite{tian2020polarizability}, is large enough to compete with the
	interlayer hybridization of the low energy states (Fig. S6).
	
In conclusion, we find that a displacement field reconstructs the low-energy
band structure of six-layer PtSe$_2$. Fixed-relaxation-time Wannier transport
calculations attribute the conductance enhancement near charge neutrality to
displacement-field-induced band overlap and the suppressed conductance at high
hole density to valence-band reconstruction. The evolution of the on-state
conductance therefore tracks the field-induced evolution of the conducting
bands, revealing band reconstruction at finite carrier density.
	
\begin{acknowledgments}
	A.A. discloses support for the research of this work from the National Research Foundation, Prime Minister's Office, Singapore (NRFF14-2022-0083) and Ministry of Education--Singapore (23-0683-A0001). Y.Z., J.F., and O.V.Y. acknowledge support from the Swiss National Science Foundation (Grant Nos.~204254 and 224624). First-principles calculations were performed at the Swiss National Supercomputing Centre (CSCS) under Project No.~lp96 and using the facilities of the Scientific IT and Application Support Center of EPFL. Z.S. discloses support from the ERC-CZ program (Project LL2101) of the Ministry of Education, Youth and Sports (MEYS) and from the project Advanced Functional Nanorobots (Reg. No.~CZ.02.1.01/0.0/0.0/15\_003/0000444), financed by the EFRR. K.W. and T.T. disclose support from JSPS KAKENHI (Grant Nos.~21H05233 and 23H02052) and the World Premier International Research Center Initiative (WPI), MEXT, Japan.
\end{acknowledgments}

    \nocite{weintrub2022intense,prakash2017bandgap}
	\bibliographystyle{apsrev4-2}
	\bibliography{references}
	
	\section*{APPENDIX A: GROWTH OF CRYSTALS}
	
	PtSe\textsubscript{2} single crystals were synthesized from platinum powder
	(99.99\%, SurePure) and selenium granules (99.9999\%, 2--4\,mm, Wuhan Xinrong
	New Materials Co.) using a self-flux method in evacuated quartz ampoules.
	Stoichiometric amounts corresponding to 3\,g of PtSe\textsubscript{2}, together
	with 2\,at.\% excess selenium, were sealed under high vacuum
	(\textless{}\(1 \times 10^{-3}\) Pa) and enclosed in a corundum crucible. The
	ampoules were heated to 800\,$^\circ$C at 1\,$^\circ$C\,min$^{-1}$ and held for
	25\,h, followed by heating to 1280\,$^\circ$C, slow cooling to
	1230\,$^\circ$C, and further cooling to 1000\,$^\circ$C. The samples were then
	furnace-cooled to room temperature. The resulting crystals were recovered in
	an argon-filled glovebox and mechanically exfoliated for subsequent
	measurements.
	
	Hexagonal boron nitride (hBN) single crystals were grown using a
	temperature-gradient method under high-pressure and high-temperature conditions
	(3\,GPa, 1500\,$^\circ$C for 120\,hours), with a Ba--BN solvent to achieve high
	purity. The as-grown crystals were heat treated at 2000\,$^\circ$C under
	nitrogen to reduce oxygen impurities, followed by chemical cleaning in hot aqua
	regia to remove residual solvent, as reported previously.
	
	\section*{APPENDIX B: DEVICE FABRICATION AND TRANSPORT MEASUREMENTS}
	
	PtSe\textsubscript{2}, graphite, and hBN flakes were mechanically exfoliated
	from bulk crystals onto Si/SiO\textsubscript{2} (300 nm) substrates. Graphite
	bottom gates were assembled using a PDMS/PC-assisted dry-transfer process, with
	graphite flakes aligned beneath clean regions of the bottom hBN. Contacts were
	defined by standard electron-beam lithography, followed by reactive ion etching
	of the exposed regions to a depth of approximately 15 nm. Pre-patterned
	contacts were fabricated by electron-beam evaporation of Ti/Pt (2/14 nm). The
	devices were then annealed under high vacuum
	(\textasciitilde1 $\times$ 10$^{-6}$ mbar) at 340 $^\circ$C for 6 hours to
	remove fabrication residues. Subsequently, top hBN and
	PtSe\textsubscript{2} flakes were transferred onto the pre-patterned Pt
	contacts. Final electrical contacts (Ti/Au, 2/100 nm) were defined by a second
	lithography and evaporation step, followed by a final annealing treatment under
	the same conditions to further clean the heterostructure interfaces.
	
	Electrical transport measurements were performed in a closed-cycle
	\textsuperscript{4}He cryostat (Oxford TeslatronPT). Gate voltages were applied
	using a home-made digital-to-analog converter system, and two-terminal DC
	transport measurements were carried out using Keithley 2450 SourceMeters.
	
	\section*{APPENDIX C: WANNIER TIGHT-BINDING MODEL CONDUCTANCE}
	
	We performed a first-principles calculation of the electronic structure of
	six-layer PtSe\textsubscript{2} using the GPAW code~\cite{enkovaara2010gpaw}
	within LDA ~\cite{perdew1981sic}, with a plane-wave cutoff of 500 eV and a
	12 $\times$ 12 $\times$ 1 Brillouin-zone mesh. The atomic structure was
	relaxed until the residual forces were below 1 meV/\AA{}. A 0.75 eV
	scissor-operator correction was included in the self-consistent calculation by
	shifting the unoccupied states. We used this correction since DFT does not
	reproduce the PtSe\textsubscript{2} gap quantitatively. Calculations beyond
	DFT, including G$_0$W$_0$ studies of monolayer PtSe\textsubscript{2}
	~\cite{zhuang2013photocatalysts,sajjad2018excitons}, together with
	spectroscopic and transport measurements on ultrathin
	PtSe\textsubscript{2}~\cite{zhang2021precise,ciarrocchi2018ptse2,
		wang2015monolayer}, show a strong dependence of the gap on layer number.  The scissor shift sets the band separation used to construct the Wannier tight-binding model
	~\cite{marzari1997mlwf,souza2001entangled} while preserving the orbital
	character and in-plane dispersion of the bands. The Wannier projections were
	Pt-d and Se-p orbitals on each PtSe\textsubscript{2} layer, following the
	maximally localized Wannier construction as implemented in Wannier90
	~\cite{mostofi2014wannier90,pizzi2020wannier90}. For six layers, this gives 66
	orbital Wannier functions before spin-orbit coupling. The transport calculation
	was performed with the spinor Wannier Hamiltonian, which contains 132 Wannier
	functions in the final model. This basis keeps the layer and orbital character
	of the low energy states explicit and allows the conductance to be evaluated on
	dense momentum grids by Wannier interpolation ~\cite{yates2007interpolation}
	and WannierBerri ~\cite{tsirkin2021wannierberri}.
	
	We applied the out-of-plane electric field at fixed atomic geometry. For each
	value of the field, the Hamiltonian was written as
	
	\begin{equation}
		H(E_z) = H_0 + E_z \widehat{z},
		\label{eq:field-hamiltonian}
	\end{equation}
	
	where $\widehat{z}$ is the out-of-plane position operator measured from the
	middle of the slab. For each \emph{E}\textsubscript{z}, the Fermi level was
	obtained from the integrated density of states, and the final maps were
	parameterized by carrier density relative to charge neutrality. We evaluated
	the longitudinal conductance in the constant-relaxation-time approximation. We used $\tau$ = 100 fs throughout. The absolute value of
	\emph{G} scales linearly with this relaxation time, while the dependence on
	electric field and carrier density is determined by the Wannier tight-binding model band structure.
	The calculation therefore describes the band structure contribution to the
	conductance and does not include contact resistance or changes in disorder
	scattering.
	
\end{document}